\documentclass[fleqn,usenatbib]{rasti}

\usepackage{newtxtext,newtxmath,hyperref}
\usepackage{upgreek}
\usepackage[usenames,dvipsnames]{xcolor}

\definecolor{chgcol}{rgb}{1.0, 0.44, 0.37}
\newcommand{\chg}[1]{{#1}}
\usepackage[T1]{fontenc}

\usepackage{graphicx}	
\usepackage{amsmath}	

\title[Audification of Galaxy Spectra]{Inspecting spectra with sound: proof-of-concept \& extension to datacubes} 

\author[James W. Trayford et al.]{
James
 W. Trayford,$^{1}$\thanks{E-mail: james.trayford@port.ac.uk}
C.M. Harrison,$^{2}$\thanks{E-mail: christopher.harrison@newcastle.ac.uk}
R.C. Hinz,$^{2}$
M. Kavanagh Blatt,$^{1}$   
S. Dougherty,$^{2}$ \&
A. Girdhar$^{2,3}$
\\
$^{1}$Institute of Cosmology and Gravitation, University of Portsmouth, Dennis Sciama Building, Burnaby Road, Portsmouth PO1 3FX, UK \\
$^{2}$School of Mathematics, Statistics and Physics, Newcastle University, NE1 7RU, UK\\
$^{3}$Ludwig-Maximilians-Universit{\"a}t, Professor-Huber-Platz 2, D-80539 M{\"u}nchen, Germany\\}

\date{Accepted XXX. Received YYY; in original form ZZZ}

\pubyear{2023}

\begin{document}
\label{firstpage}
\pagerange{\pageref{firstpage}--\pageref{lastpage}}
\maketitle

\begin{abstract}
We present a novel approach to inspecting galaxy spectra using sound, via their direct audio representation (\textit{`spectral audification'}). We discuss the potential of this as a complement to (or stand-in for) visual approaches. 
We surveyed 58 respondents who use the audio representation alone to rate 30 optical galaxy spectra with strong emission lines. Across three tests, each focusing on different quantities measured from the spectra (signal-to-noise ratio, emission-line width, \& flux ratios), we find that user ratings are well correlated with measured quantities. This demonstrates that physical information can be independently gleaned from \textit{listening} to spectral audifications.
We note the importance of context when rating these sonifications, where the order examples are heard can influence responses. Finally, we adapt the method used in this promising pilot study to spectral datacubes. We suggest that audification allows efficient exploration of complex, spatially-resolved spectral data.
\end{abstract}




\section{Introduction}
Sonification is the process of conveying data using non-verbal sound \citep[][]{Kramer1999}. Sound has the potential to provide unique advantages when inspecting data. For example, the ear is particularly good at parsing evolving or transient signals, with a \chg{larger dynamic range in frequency than the eye \citep{Heffner2007, Sliney2016}}. Audio also provides an additional modality to inspect multi-dimensional data, mitigating the limits of purely visual methods. This is promising in the \textit{‘big data’} era of astronomy, where innovation is needed to explore complex, multidimensional data. Indeed, sonification has long existed, finding application in science and engineering \cite[][]{Hermann2011}, and seeing a surge of interest in astronomy in recent years \citep[][]{Zanella2022}. 

Sound has become established in astronomy as a means to provide  educational resources and  accessible channels for visually impaired (VI) individuals, \chg{either in pure audio or multi-modal context, often by using sound to accompany and enhance visualisations} \citep[][]{Alexander2011,Harrison2022AG,Bardelli2022,GarciaBenito2022,Zanella2022}. 
However, work is on going to demonstrate the utility of sonification as a research tool. 
\citet{DiazMerced2013} and \citet{TuckerBrown22} provide precedent, demonstrating how signals can be discerned from sonified lightcurves in both multimodal (sound \& visual) and standalone contexts, through controlled user tests. 

In this pilot study, we further explore the research utility of sonification. Particularly, we prototype applications to \textit{(hyper-)spectral datacubes}, a form of data that is transforming observational astronomy with the rise of resolved \textit{spectroscopy} and \textit{interferometry} \citep[e.g.][]{Bacon2010, Boker2022}, and across other fields from biomedical engineering to geophysics \citep[e.g.][]{Cassidy2004, Bernhardt2007, Li2013}. While the detail and dimensionality (i.e. 3D) of these datacubes makes purely visual inspection difficult, we posit that the spectral dimension can naturally be represented with sound, accompanying the two spatial dimensions typically projected for visual display. This could enable efficient datacube exploration \& build intuition for otherwise unwieldy data. 

We utilise the {\tt STRAUSS}\footnote{\textbf{S}onification \textbf{T}ools and \textbf{R}esources for \textbf{A}stronomers \textbf{U}sing \textbf{S}ound \textbf{S}ynthesis.} code to  sonify each galaxy spectrum in our experiments \citep[][]{STRAUSS}. {\tt STRAUSS} is a modular python package providing diverse sonification options for data analysis and presentation, allowing parameters of the data to be directly mapped to expressive properties of sound (e.g. pitch, stereo field placement, timbre, etc), via synthesis or manipulation of recorded sounds, or `samples' \citep[applied in][]{Harrison2022AG,TuckerBrown22}.

In astronomy-associated tools for sonifying 1D data, i.e. where  $y= f(x)$, typically the $y$-variable (e.g. brightness) is mapped to a sound property such as pitch, and its variation with the $x$-variable (e.g. date) is represented by the modulation of a carrier tone over time (see e.g.   {\tt astronify}\footnote{\href{https://astronify.readthedocs.io/en/latest/}{\tt astronify.readthedocs.io}}). However, a limitation of the parameter modulation approach is the need to `\textit{play out}' the entire $x$-domain to fully inspect the data. Playback must therefore be slow enough for subtle or small-scale variations in the data to be perceptible. This is particularly pertinent for narrow, highly informative features, like spectral lines. Audio inspection of large data sets (or even a single modern datacube) can therefore be a slow process. An alternative is \textit{Spectral Audification} \citep[SA,][]{Sturm05,Newbold15}, harnessing the analogy between electromagnetic and audio spectra. By scaling the flux as a function of electromagnetic wavelength to an audible frequency range, an audio signal can be  synthesised, representing this distribution of relative power across perceptible frequencies. This provides an `instantaneous'
representation of the spectrum where various frequency components can be heard simultaneously. \chg{This can be considered a form of \textit{`spectral compression'}}.

\chg{\chg{We note that despite it's direct or intuitive relation to the data, such \textit{`audification'} isn't always necessarily the most effective, with more abstracted sonified representations sometimes superior at conveying data \citep[e.g.][]{Franinovic2013}}. Data sonification innovations have emerged across many data fields, both in analytic contexts, as well as for more outreach-based or artistic applications. Sonification of multidimensional medical data for diagnostic purposes exemplify analytic applications, including electroencephalographic data \citep{Valjamae2013} and optical coherence tomography data \citep{Ahmad2010}. Audio representations of spectral data have been demonstrated through a similar SA approach (e.g. nuclear magnetic resonances \citealt{Morawitz2019}, X-ray scattering data \citealt{Schedel2012}, the CMB angular power spectrum \citealt{McGee2011}) but also abstractly as e.g.
 feature sets via autoencoder training on stellar spectral libraries \citep{Riber2023}. Such approaches may find use in more analytic or quantitative contexts. Here, we focus on a spectral audification approach in particular.}

We propose this SA approach as a promising method (in both multimodal and standalone contexts) for inspecting spectroscopic data, particularly in the context of spectral datacubes. In this study, we sonify individual galaxy spectra representing a range of physical properties, and evaluate how well their relative properties can be gleaned from SA sonification via a series of rating tasks, comprising a survey. \S~\ref{sec:sonification} details the sonification approach, \S~\ref{sec:methods} details the spectra selection and survey design, \S~\ref{sec:results} then presents the results from the 58 respondents (\ref{sec:ratings}), and the broader applications to spectral datacubes (\ref{sec:ifu}). We conclude in \S~\ref{sec:conclusion}. For clarity, we refer to light spectra using their \textit{wavelength}, and sound using \textit{frequency}.

\section{Spectral Audification}
\label{sec:sonification}
We synthesise the audio representing each spectrum \textit{additively}, with a superposition of sinusoidal waves.
A sine wave represents each wavelength, $\lambda_{i}$, of the galaxy spectra, with amplitude corresponding to the observed flux density,  $F_{i}$. These waves are summed, following
\begin{equation}
A(t) = \sum_{i} F_{i}\sin\left(2\uppi 
\nu_{i}t+\phi_{i}\right),
\end{equation}
where $t$ is time in seconds, $\phi_{i}$ is a $[0,2\uppi]$ uniformly-distributed random phase (important to avoid imposing spurious phase correlations), and $\nu_{i}$ is the audio frequency mapped from each wavelength of the galaxy spectrum, with
\begin{equation}
\nu_{i} = \frac{\lambda^{-}\left(\lambda^{+} - \lambda_i\right) \left(\nu^{+} -\nu^{-} \right)}
{\lambda_i\left(\lambda^{+} - \lambda^{-}\right)}
 + \nu^{-},
\end{equation}

where $+$ and $-$ superscripts indicate maxima and minima of value ranges. We chose a minimum audio frequency of $\nu^{-}$ = 250\,Hz with a maximum of $\nu^{+}$ = 1250\,Hz. This range was motivated by: (1) a range typical human hearing has an almost constant level of frequency resolution \citep[i.e. the ability to audibly distinguish between two different frequencies,][]{Zwicker1957} and; (2) there is not dramatic changes in \chg{\textit{perceived}} loudness \chg{at a fixed pressure level} across the chosen range \chg{for pure (i.e. monochromatic) tones} \citep[][]{Suzuki2004}. \chg{Our complex tones, could exhibit differing behaviour, however this is taken as a useful guide}. These considerations also helped to ensure the sounds were in a comfortable frequency range, avoiding the highest frequencies that can be particularly uncomfortable or tiring to listen to. We produced SAs of 3\,s duration, allowing participants to replay the audio (see \S~\ref{sec:survey}).

We note that this additive synthesis approach affords maximal flexibility, allowing arbitrary wavelength ranges and spacings to be represented, particularly useful for the explorative prototyping of this pilot study. However this approach is typically much slower than using an Inverse Fast Fourier Transform (IFFT). In \S~\ref{sec:ifu} we discuss employing IFFT to produce SAs orders-of-magnitude faster.

\section{Participant testing and spectra used for spectral audification}
\label{sec:methods}

Our study was designed to assess if participants ranked spectra on what they {\em heard} from hearing a SA of the spectra, then these ratings would correspond to the emission-line properties. Across three separate tests, detailed below, we assessed the participants' ability to hear variations in three emission-line properties: (a) signal-to-noise ratio (SNR); (b) emission-line width and; (c) emission-line flux ratios. The survey design is presented in \S~\ref{sec:survey} and the galaxy spectra used for each test is presented in \S~\ref{sec:spectra}. A transcript of the survey form and the sounds presented to participants are made available\footnote{\chg{See \href{https://doi.org/10.25405/data.ncl.22816442}{doi.org/10.25405/data.ncl.22816442}}}.

\begin{figure*}
\includegraphics[width=0.9\textwidth]{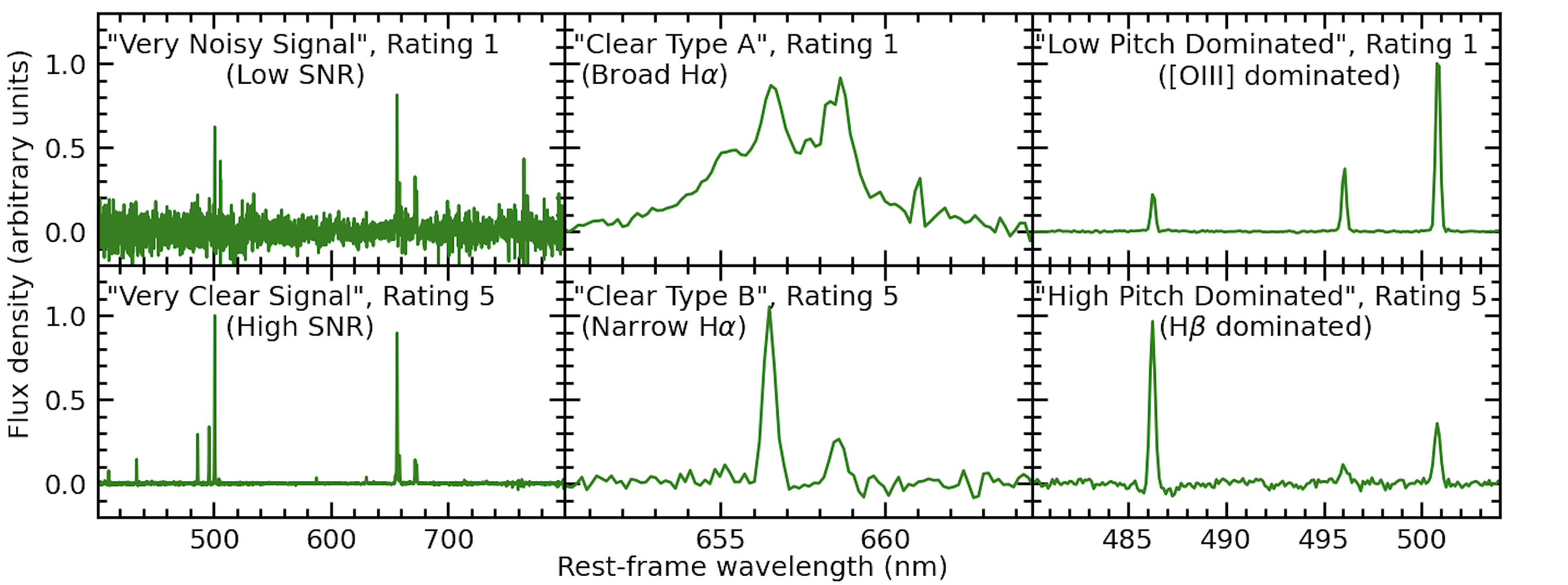}
\caption{Visual representations of galaxy spectra that were played as audio files to the participants during the initial training (the visuals were not shown). The three columns correspond to the three tests (SNR, emission-line width, and emission-line ratio). In each case, the top panel shows an example of Rating 1 and the bottom, an example of Rating 5. The audio files can be found in the supplementary material.}
    \label{fig:examples}
\end{figure*}

\subsection{Survey design}
\label{sec:survey}
We performed an anonymous participant survey 
via \textit{Google Forms}, using embedded video (with blank visuals) to play sounds. Participants were not informed that the sounds corresponded to galaxy spectra, but simply that they survey was part of {\em ``exploring new ways to represent astronomy with sound.''}  
We also collected data on demographics  (musical and data analysis experience, optionally age) and user experience, but forego its analysis in this pilot study, given our limited sample size. 

We ran three separate tests, each testing variation in a different quantity of interest (QoI, see \S~\ref{sec:spectra}).
At the beginning of each of the tests, six example sounds were presented to participants, functioning as a tutorial: three examples of the very low QoI values ('Rating 1') and three examples of very high values ('Rating 5'). For each test, the participants were then asked to rate a set of 10 sounds (presented in pseudo-random order) with an integer value of 1--5, inclusive. Of these 10 sounds, one was a duplicate (i.e., each test contained 9 unique sounds, of which 1 was presented to the participants twice), where we avoid duplicates being presented consecutively. The survey was run from 8th -- 29th March 2022, with 58 responses. \chg{Of these respondents, 26 were aged $<$~30, 19 aged 30-50 and 10 aged $>$~50. Users also exhibited diversity in their self-assessed experience as musicians (data analysts) with 20 (37) \textit{Experienced} or \textit{Very Experienced}, 30 (14) with some level of experience and 8 (7) with \textit{No Experience}.}

\subsection{Galaxy spectra}
\label{sec:spectra}
We opted to use spectra of \textit{real} galaxies (as opposed to synthetic or idealised data). All spectra and corresponding automated emission-line properties, are taken from the Sloan Digital Sky Survey (SDSS) Data Release 7 \citep[][]{Abazajian09}. We focused on emission-line properties, using continuum-subtracted spectra. Additionally, we select a narrow redshift range of $z=0.09$--0.11 and required a signal-to-noise ratio (SNR) of $>$3 for each  emission line of interest (H$\beta$, [O~{\sc iii}], H$\alpha$, and [N~{\sc ii}]). We shifted all spectra to the rest-frame, using the spectroscopic redshifts provided by SDSS.

{\bf Test A: Variations in SNR:} For this sub-sample we were interested if participants could successfully rate the SNR of spectral emission lines from their SAs alone. To minimise other factors (than SNR) changing the sound, we excluded Type 1 active galactic nuclei (AGN) and used a relatively narrow emission-line flux ratio selection, with $\log_{10}({\rm [O~III]/H\beta}) \in [0.3,0.6]$ and $\log_{10}({\rm [N~II]/H\alpha}) \in [-0.7,-0.4]$.
 We calculated an average, indicative, SNR for each spectra by taking an average over the four emission lines: H$\beta$, [O~{\sc iii}], H$\alpha$, and [N~{\sc ii}]. For the six examples presented during the participant training, we used three galaxy spectra with low SNRs of 9--13, which we described as `Very Noisy Signal' (Rating 1) and three galaxy spectra with high SNRs of 84--122, described as `Very Clear Signal' (Rating 5, see first column of Fig.~\ref{fig:examples}). For the nine cases used in  participant testing, we randomly selected spectra covering a SNR range 11--115. The wavelength range used for their sonification was 400-800\,nm, presented in the left column of Fig.~\ref{fig:examples}.  
 
{\bf Test B: Variations in emission-line width:} For this sub-sample we were interested if participants could successfully rate the H$\alpha$ emission-line width of spectra, from SA alone. We used a wavelength range of 650--665\,nm (i.e., centred around H$\alpha$ and [N~{\sc ii}]), 
presented in the central column of Fig.~\ref{fig:examples}. To minimise the influence of varying emission-line flux ratios, we only used spectra in a narrow range of $\log_{10}({\rm [N~II]/H\alpha}) \in [-0.7,-0.4]$. 
For the six examples presented during the participant training, we used three spectra with low values of full-width at half maximum (FWHM) for H$\alpha$ (FWHM = 220--240\,km\,s$^{-1}$), which we label as `Type B' (Rating 5) and three examples with large widths of H$\alpha$ FWHM=2400--2900\,km\,s$^{-1}$, labelled as `Type A' (Rating 1). We note that this labelling was chosen in lieu of clear shared vocabulary to describe the distinct qualities of the sounds. This results in a \textit{dichotimised} rating system  for this particular test, where 1 and 5 representing `high confidence' of Type A and B, respectively, with 3 indeterminate (revisited in \S~\ref{sec:results}).

{\bf Test C: Variations in emission-line flux ratio:} For this sub-sample we were interested if participants could successfully rate [O~{\sc iii}]/H$\beta$  emission line flux ratio  in spectra, from their SA alone. To minimise the influence of emission-line width variations on the sound, we excluded Type 1 AGN and required that the FWHM of the [O~{\sc iii}] line be less than 250 km\,s$^{-1}$. 
We also required the H$\beta$ SNR to be $>$50. For the training examples, we used three galaxy spectra with high [O~{\sc iii}]/H$\beta$ ratios ($\log$[O~{\sc iii}]/H$\beta\in [0.69,0.79]$), described as `Low Pitch Dominated' (Rating 1) and three examples with low ratios ($\log$[O~{\sc iii}]/H$\beta\in [-0.48,-0.27]$), which we described as `High Pitch Dominated' (Rating 5). The wavelength range used in the sonification was 480--504\,nm, with examples shown in the right column of Fig.~\ref{fig:examples}.

\begin{figure*}
	\includegraphics[width=0.92\textwidth]{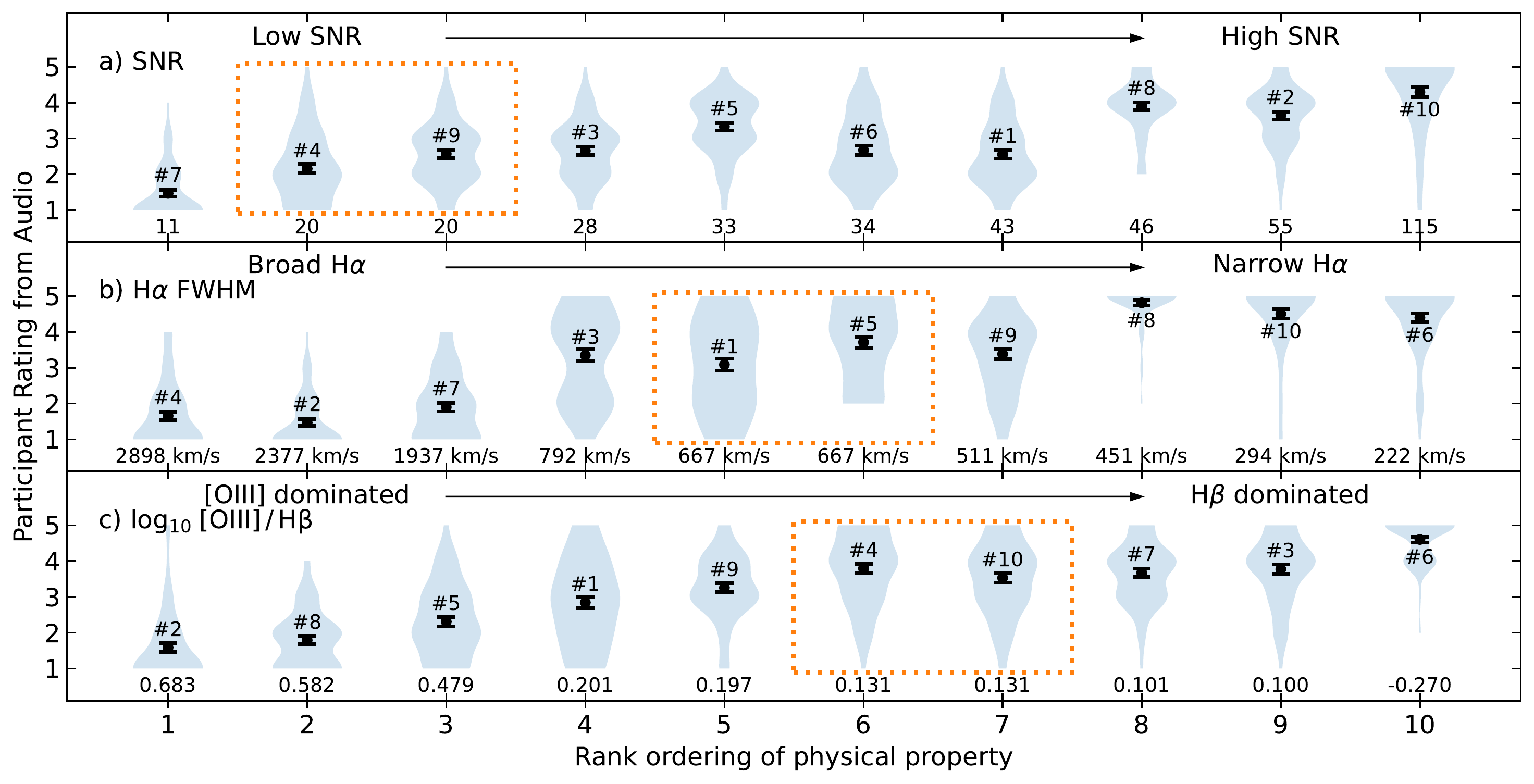}     
 \vspace{-0.35cm}
    \caption{Mean participant rating (error bars show standard error on the mean) for each of the galaxy SAs, as a function of the rank ordering of the principle physical quantify varied in each test. Panels correspond to different test ({\em top:} SNR; {\em middle:} emission-line width; {\em bottom:} emission-line ratio\chg{)}. The physical quantity is indicated in the top-left of each panel, with precise values inset along the $x$-axis (with units where appropriate). The background violin plots represent the full distribution of participant rankings. \textit{Orange dotted lines} highlight the pair of identical examples for each experiment. As their values are identical,  relative ranking instead represents the order they appear in the survey (earlier to later). The text annotating data points indicates their \textit{presentation order} to respondents.}
    \label{fig:violin}
\end{figure*}

\section{Results and Discussion}
\label{sec:results}

\subsection{Rating results}
\label{sec:ratings}

Our primary survey results are summarised in Fig.~\ref{fig:violin}, with panels a) to c) (top to bottom) corresponding to the three tests detailed in \S~\ref{sec:spectra}. The mean participant rating for each galaxy SA is plotted in rank order by QoI, such that user rating and physical ranking should \textit{positively correlate} if ratings are representative. In all three tests, the average ratings show a clear positive correlation, particularly between the extrema. In all cases the highest and lowest ranked examples are rated  as such by the plurality of users. We regard this outcome as a \textit{very encouraging} sign for the technique, particularly given our rudimentary rating tasks over just 58 respondents with limited training and knowledge about how the SA approach works.

Given the nascent stage of this approach, we wish to avoid over-interpretation of these first results. However, we perform some basic numerical analysis, focusing on rank correlation (monotonicity) for its insensitivity to the property spacing and perceived scaling of the QoI from the SAs. We also exclude the repeated SAs (right-hand distributions in  orange boxes) from this correlation for now, as their relative rank is unrelated to the QoI. The mean ratings correlate well with physical rank for all tests, with Spearman's rank coefficients of 0.86, 0.9 and 0.95 for tests a) to c), respectively. When applied to individual ratings (constituting the underlying distributions in Fig.~\ref{fig:violin}), these correlations are weaker but still positive at 0.61, 0.72 and 0.69, respectively. However, rank correlation of individual responses are affected by the limited sample size and heavy quantisation of ratings and rankings. To quantify this uncertainty, we compute  the rank correlation coefficient bounds for quantised data, proposed by \citealt{Couso2018} (based on Kendall's $\tau$, a similar metric to Spearman's). We recover [0.13, 0.71], [0.41, 0.79] and [0.2, 0.76] for tests a to c, respectively. Definitively positive correlations, but indicating that correlation \textit{strength} is poorly constrained given survey limitations. 

Clearly there is some underlying user-user variance to investigate further. For example, while the line ratio test (c) is most accurate in terms of average rating, it is less precise than the line width test (b), where the three highest and lowest rank extrema are reliably rated at high and low extremes by users, respectively. This could partly be due to how questions are posed; test c) asks users to assess the relative strength of high or low pitches in the sound, while test b) asks their confidence in sorting examples between Types A and B (broad to narrow H$\alpha$). Test b) respondents correctly rate the 3 broadest and narrowest H$\alpha$ galaxies with low and high ratings, respectively. The 4 intermediate H$\alpha$ galaxies are found to be relatively indeterminate. We suggest respondents are treating test b) as more of a sorting task, yielding more \textit{`stepped'} mean ratings corresponding to A, B and indeterminate types, contrasting the continuous variation for test c). 

Another aspect that may influence respondents is the (pseudo-random) order SAs are presented in. As a simple test, we rank examples by mean user rating (including repeats), and measure how these ranks are offset from their physical property ranking. We then compare this offset to the difference in physical rank between a SA from the one heard directly before. We find that these quantities show a significant (2$\sigma$)  rank correlation for test a) - users are more likely to (under-) over-rate the clarity of the signal if they heard a relatively (clear) noisy example before. Test b) only correlates at the 1$\sigma$ level, and test c) shows no evidence of correlation\footnote{For test c), concordance in the ordering by physical rank and mean rating (i.e. their good correlation) means there is insufficient data to measure this.}. These findings are reflected by the repeat examples; for tests a) and b), respondents are more likely to rate a repeat example \textit{higher} if the preceding SA is ranked relatively \textit{lower}, excepting test c (which incidentally shows the smallest absolute difference between the mean rating of repeats). While more investigation is needed, it is evident that \textit{context} can influence how users perceive SAs, with users often sensitive to \textit{contrast}, reacting against the SA they heard before. This sensitivity may have advantages, e.g. when inspecting continuously evolving sound or where differential changes are interesting. This is particularly relevant to our prototype spectral datacube `listener', next. \chg{Particular sensitivity to subtle \textit{relative} differences in auditory stimuli has been established in psychoacoustics,  conceptualised as a \textit{`just noticeable difference'} \citep{Zwicker2007, Ahmad2010}}.

\chg{Respondents also rated the tolerability of the SA examples, from 1 (\textit{Very intolerable}) to 5 (\textit{Tolerable}). These ratings were chosen 1, 7, 11, 26, and 13 times respectively, suggesting users generally found these on the tolerable side, though there was a minority finding the sounds difficult to listen to.}

\subsection{Future applications to datacube exploration}
\label{sec:ifu}

\begin{figure}
\centering
 \includegraphics[width=0.8\columnwidth]{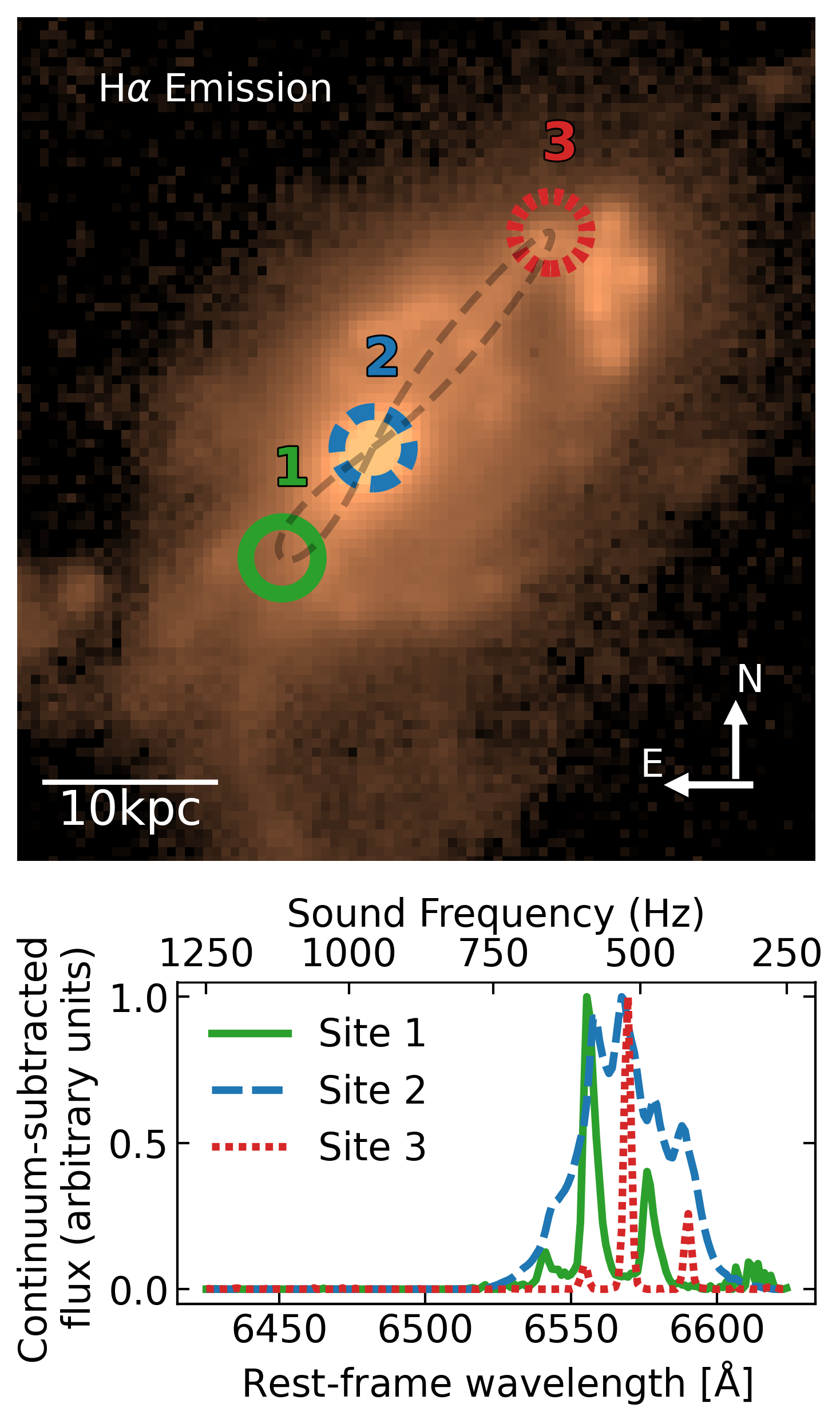} \vspace{-0.3
cm}
\caption{Prototype audification   of a spectral datacube. \textit{Top panel} shows  H${\rm \alpha}$ emisson, overlaid with three aperture positions along a figure-eight track through the galaxy center (\textit{grey dashed}). \textit{Bottom panel} shows the continuum-subtracted and normalised flux for each aperture position, over a  wavelength range capturing the H$\alpha$ line in the overall galaxy rest-frame. The upper $x$-axis illustrate how this spectrum is directly mapped to audio frequency. A full sonified animation can be found at \chg{\url{https://doi.org/10.25405/data.ncl.22816442}.}}
    \label{fig:cube}
\end{figure}

We illustrate our prototype datacube explorer in Fig.~\ref{fig:cube} (linking to sonified animation in caption), using a MUSE  \citep{Bacon2010} datacube of galaxy J1316+1753, a $z=0.15$ quasar host with defined spiral arms \cite[studied in][]{Girdhar2022}.  Here, the SA is generated from the aggregate spectrum in an aperture, freely moved around the 2 projected spatial dimensions of the datacube. We render a figure-eight path about the galaxy center, sonifying the wavelength region around the rest-frame H$\alpha$ complex. Doppler shifted emission can be heard as a distinct pitch-shift between the emission either side of the disc (sites 1 to 3). In the central region (site 2) the dominance of broad lines leads to a \textit{`noisier'} tone. Further detail can be heard, e.g. the narrow  emission and higher recession velocities around site 3. 

Ideally, a spectral datacube SA tool would be `real-time', allowing aperture size, and wavelength parameters to be adjusted on the fly. This requires low-latency SA generation. To this end we have implemented a dedicated \textit{`spectraliser'} generator type in \texttt{STRAUSS}, via an IFFT algorithm\footnote{python notebooks using this new \textit{spectraliser} approach in \texttt{STRAUSS} can be found at \href{https://github.com/james-trayford/AudibleUniverseWorkbooks/tree/group3}{github.com/james-trayford/AudibleUniverseWorkbooks/tree/group3}}. To generate the 3~s clips of our study (which could be looped and faded between based on position), this approach takes an average of 2.3~ms, 4 orders of magnitude faster than the additive approach, and crucially, safely faster than the maximum tolerable latency for interactive applications \citep{Lago2004}. This augurs well for real-time, responsive SA of spectral datacubes in the future\footnote{\chg{Such a tool would complement existing experiments and tools for resolved astronomical data, see e.g. the \textit{Soniverse} and \textit{VoxMagellan} projects}}.   

\section{Conclusions \& Future Directions}
\label{sec:conclusion}

We present spectral audification (SA) as a novel, auditory approach for inspecting astronomical spectra, to complement and enhance traditionally visual approaches in research, while also providing an accessible channel for VI people to inspect data. Using galaxy spectra (Fig.~\ref{fig:examples}), we demonstrate how key properties (e.g. SNR, linewidth, line ratios) can be gleaned from their audio representation alone, via a 58-respondent survey. We show that user ratings correlate appropriately with their physical ranking in Fig.~\ref{fig:violin}; very encouraging in the context of our small sample and minimal training of users.  User-user variance is noted, and partly attributed to the role of \textit{context} (SA order), finding that mean user ranking offsets can correlate with how the QoI has changed from the spectrum heard immediately prior (particularly for SNR, \S~\ref{sec:survey}). We note that this sensitivity to \textit{contrast} may be useful in the context of our prototype datacube application. We provide an illustration and sonified animation of a prototype SA explorer for spectral datacubes (Fig.~\ref{fig:cube}), noting that live sonification of spectral datacubes are feasible given our modified IFFT approach.

In future surveys, improving our perceptual evaluation methods \citep{Misdariis2022}, and larger sample sizes could give quantitative insight into the reliability of SA inspection, beyond our proof-of-concept (see \S~\ref{sec:ratings}). The role of context could probed further, for example, by randomising SA order by user (as e.g. a \textit{Zooniverse} project). \citet{TuckerBrown22} note the importance of training when introducing sonification approaches for research. In this work, we provided limited training, also deliberately giving users no insight or explanation of the technique (\S\ref{sec:survey}). Complementary, extended evaluation studies where users can develop more knowledge and experience around the approach (particularly for the datacube explorer) may better reflect the potential of audio modalities; after all, most researcher's visual inspection skills are honed over decades. 

\chg{More extensive testing is also important to understand how the training rate and overall effectiveness of SA inspection may vary statistically with demographic factors, including age, experience and level of vision. Similarly, comparisons between the effectiveness of visual, auditory and hybrid inspection modes (i.e. additional testing including visual representations of spectra in 1D or as spectrograms) could be informative. } \chg{Developing effective training approaches and materials for SA (as e.g. \citealp{Alexander2011}), alongside improved sonification techniques, appears key to better outcomes from auditory inspection of data. We  also identify a number of avenues for  investigation to improve the aesthetics of future SA representations.}

In the short-term, we will work on the spectral datacube explorer (\S\ref{sec:ifu}), seeing great potential for data intuition, accessibility and discovery. With the proliferation of (hyper)spectral datacubes, enhanced inspection modes could improve efficiency in research. The discovery potential of SA, where certain aspects may be clearer or enhanced through sound, could offer invaluable unforeseen benefits (e.g. finding signals or artefacts that could be otherwise overlooked in the confounding detail of modern spectral datacubes). Longer term, integration of datacube listeners into online data archives (e.g. MAST\footnote{\url{https://mast.stsci.edu/viz/ui/}}) 
is a goal. \chg{By demonstrating the efficacy of analytical sonification in astronomical data}, we \chg{could} also \chg{generalise our methods} to other fields involving spectral (datacube) analysis \citep[e.g. medicine \& engineering][]{Cassidy2004, Bernhardt2007, Li2013}, \chg{complementing prior work on audio representations and their practical efficacy in those data fields \citep[e.g.][]{Ahmad2010,
 Lee2022}}. 

\section*{Data Availability}
\chg{A survey transcript and all of the audio files used during the tutorials and testing, along with our animated prototype of a sonified IFU datacube, can be found here: \url{https://doi.org/10.25405/data.ncl.22816442}.} 

\section*{Acknowledgements}
\chg{We thank the anonymous referees for a their constructive feedback which helped to improve the manuscript, and suggesting the term `spectral compression'.} We also thank all of the anonymous, volunteer participants who carried out the surveys. We thank the Lorentz Centre for hosting the Audible Universe workshop that partly inspired this work and Scott Fleming for encouraging and practical discussion of how to integrate an audio IFU explorer into MAST. We also thank Stephen Barrass for useful discussions. \chg{JWT acknowledges support via the \textit{STFC Early Stage Research \& Development Grant}, reference ST/X004651/1.} CMH acknowledges funding from an United Kingdom Research and Innovation grant (code: MR/V022830/1).

\bibliographystyle{mnras}
\bibliography{example} 

\bsp	
\label{lastpage}
\end{document}